\documentclass[twocolumn]{aastex61}
\usepackage{graphicx}

\newcommand{\coline}{CO 2$\rightarrow$1}
\newcommand{\msun}{M$_{\odot}$}
\newcommand{\kms}{km s$^{-1}$}
\newcommand{\mics}{$\mu$m}

\begin{document}

\title{An accreting supermassive black hole irradiating molecular gas in NGC 2110}
\shorttitle{The lacuna in NGC 2110}

\author[0000-0002-0001-3587]{David J. Rosario}
\affiliation{Centre for Extragalactic Astronomy, Department of Physics, Durham University, South Road, Durham, DH1 3LE, UK}

\author{Aditya Togi}
\affiliation{St. Mary's University, San Antonio, Texas, USA}
\affiliation{University of Texas, San Antonio, Texas, USA}

\author{Leonard Burtscher}
\affiliation{Leiden Observatory, Leiden University, PO Box 9513, 2300 RA Leiden, The Netherlands}

\author{Richard I. Davies}
\affiliation{Max Planck Institut f\"ur Extraterrestriche Physik, Giessenbachstrasse 1, Garching bei M\"unchen, 85748, Germany}

\author{Thomas T. Shimizu} 
\affiliation{Max Planck Institut f\"ur Extraterrestriche Physik, Giessenbachstrasse 1, Garching bei M\"unchen, 85748, Germany}

\author{Dieter Lutz}
\affiliation{Max Planck Institut f\"ur Extraterrestriche Physik, Giessenbachstrasse 1, Garching bei M\"unchen, 85748, Germany}


\begin{abstract}

The impact of Active Galactic Nuclei (AGN) on star formation has implications for our understanding of the relationships between supermassive black holes and their galaxies, as well as for the growth of galaxies over the history of the Universe. 
We report on a high-resolution multi-phase study of the nuclear environment in the nearby Seyfert galaxy NGC 2110 using the Atacama Large Millimeter Array (ALMA), {\it Hubble} and {\it Spitzer} Space Telescopes, and the Very Large Telescope/SINFONI. We identify a region that is markedly weak in low-excitation CO $2\rightarrow1$ emission from cold molecular gas, but appears to be filled with ionised and warm molecular gas, which indicates that the AGN is directly influencing the properties of the molecular material. Using multiple molecular gas tracers, we demonstrate that, despite the lack of CO line emission, the surface densities and kinematics of molecular gas vary smoothly across the region. Our results demonstrate that the influence of an AGN on star-forming gas can be quite localized. In contrast to widely-held theoretical expectations, we find that molecular gas remains resilient to the glare of energetic AGN feedback.
\end{abstract}

\section{Introduction}

Stars form exclusively in the cold, dense molecular interstellar medium \citep[ISM;][]{kennicutt89, bigiel08}.
Active Galactic Nuclei (AGN) alter the excitation and chemistry of cold molecular gas, 
an important pathway that can suppress future star formation in galaxies \citep{sternberg94,usero04,krips08} 
and establish a co-evolutionary connection between black hole and galaxy growth 
\citep{alexander12, kormendy13, heckman14}.
Molecular spectroscopy has uncovered indirect evidence that AGN can alter central molecular gas, 
usually from the enhanced intensity of the rotational lines of the HCN and HCO+ molecules in active nuclei \citep{kohno03, usero04, krips11,
kohno08, izumi13, garciaburillo14, querejeta16, imanishi16}. However, conditions unrelated to the AGN can also boost these lines, such as
high gas densities, high molecular abundances, and infrared pumping \citep{sternberg94, izumi13, izumi16}. 

Here we present evidence for localised transformation of molecular gas through direct impact from the AGN's radiation field
in the nearby Seyfert 2 galaxy NGC 2110 (luminosity distance $D_{L} = 34$ Mpc, $cz=2335$ \kms).
In Section \ref{datasets}, we present the various high-resolution and ancillary datasets used in this work, followed
by an imaging and spectroscopic analysis, including a modeling of molecular lines, that reveals the interaction
and its properties.


\section{Observations and data preparation} \label{datasets}

Table \ref{datasettable} summarises the multi-wavelength data used for this study. Unless otherwise specified, 
we employed standard pipelines to reduce these data, adopting parameters recommended by the respective observatories.
The various images used in this work are brought together for context in Figure \ref{contextfigure}.

\subsection{ALMA spectroscopy} 

From the reduced ALMA dataset, we used CLEAN to generate a 1mm continuum map and a \coline\ cube with a velocity resolution of 5 \kms, 
both with a common restoring beam of $0\farcs71 \times 0\farcs45$ (PA of $-79^{\circ}$). 
We resampled these maps to $0\farcs24$ square spaxels for our final measurements. 

We obtained a \coline\ map by integrating the cube in velocity across a 900 \kms\ window centred on the systemic velocity of the galaxy
(Figure \ref{contextfigure}a).
We also measured \coline\ kinematics directly from the cube by fitting a single gaussian to the line in each spaxel 
with an integrated S/N $ > 5$, using the Python package \href{lmfit.github.io/lmfit-py/index.html}{LMFIT}.

\subsection{Optical and near-infrared (NIR) {\it HST} imaging} \label{hstdata}

We used narrow-band images (FR680P15) covering the H$\alpha$ and [N II]$\lambda\lambda 6548, 6584$ emission line complex
from which we scaled and subtracted an associated line-free optical broad-band image (F791W), to generate a 
pure emission line map of the circumnuclear region (Figure \ref{contextfigure}a, Figure \ref{linemaps}). 

We produced a color map (Figure \ref{contextfigure}c) by dividing the  F791W image by the deep 
NIR image (F200N). The smooth stellar light profile of NGC 2110 makes the NIR image an 
ideal backdrop for the dust features that stand out in the optical. However, the
nucleus of NGC 2110 emits continuum at 2 \mics\ from hot nuclear dust ($> 1000$ K)
that is invisible at optical wavelengths \citep{burtscher15}. This produces a nuclear red excess in the color map 
of the size of the PSF of NIR image (FWHM $\approx0\farcs26$). This region of anomalous color is disregarded
in our analysis. 

\begin{table}[t]
\center
\caption{Summary of observational datasets}
\label{datasettable}
\begin{tabular}{|l|l|l|}
\hline 
{\bf Telescope/Instrument} & {\bf Filter/Band} & {\bf Program ID} \\
\hline \hline
ALMA & Band 6 & 2012.1.00474.S \\
{\it HST}/WFPC2 & FR680P15 & 8610 \\
{\it HST}/WFPC2 & F791W & 8610 \\
{\it HST}/NICMOS (NIC3) & F200N & 7869 \\
VLT/SINFONI (AO) & K & 086.B-0484(A)\\
VLT/SINFONI (AO) & J & 060.A-9800(K) \\
{\it Spitzer}/IRS & SH+LH & AOR: 4851456 \\
\hline \hline
\end{tabular}
\end{table}

\subsection{VLT/SINFONI integral field unit (IFU) spectroscopy} 

Using a custom pipeline, we reduced both SINFONI datasets to cubes with a  
plate scale of $0\farcs05$ to take full advantage of the resolution offered by Adaptive Optics (AO).

\begin{figure*}
\includegraphics[width=\textwidth]{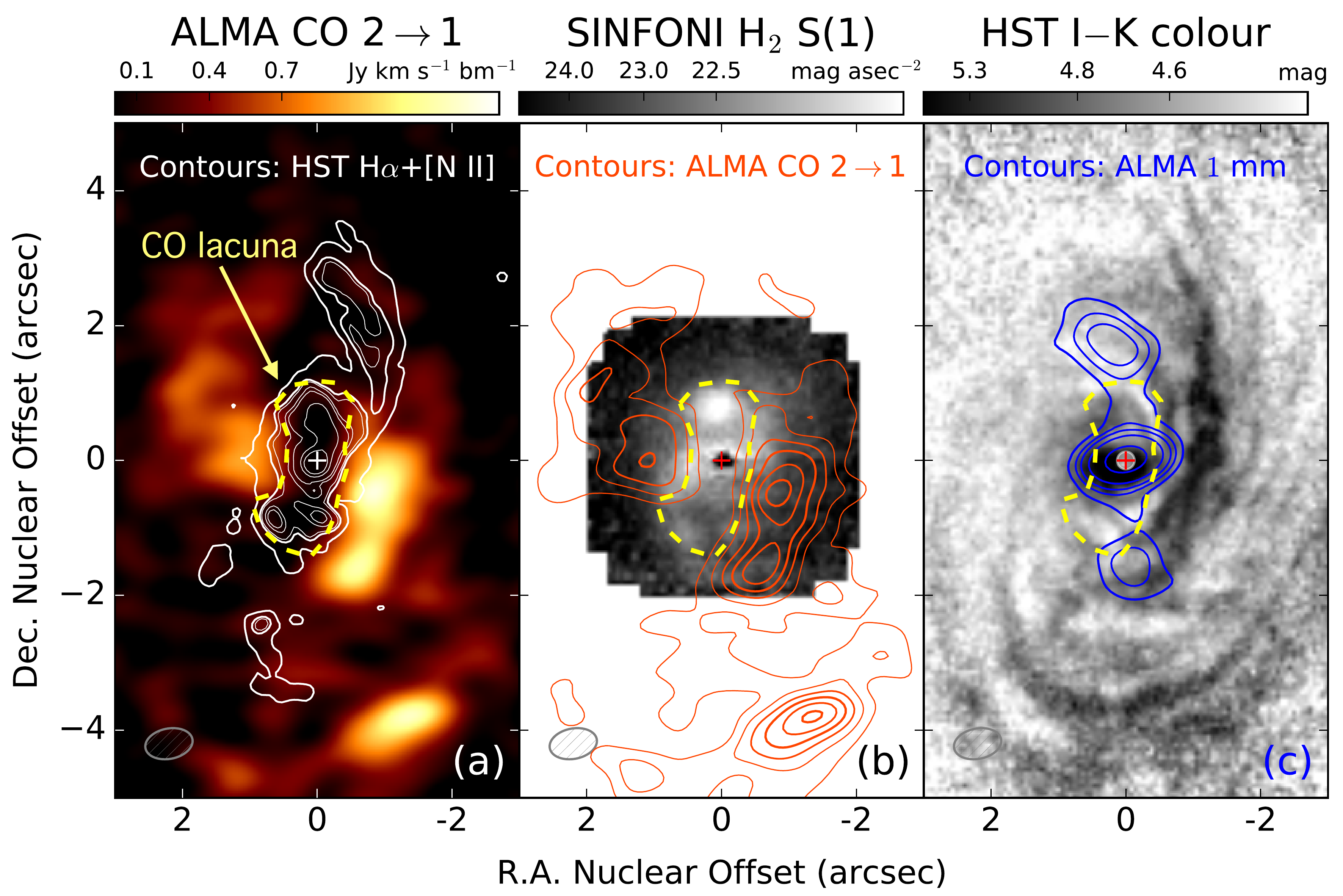}
\caption{A multi-wavelength view of the central region of NGC 2110. North is up and East to the left.
In all three panels, the nucleus is marked with a cross, the ALMA synthesised beam is shown as a grey ellipse,
and the region of CO lacuna (see Section \ref{lacuna}) is demarcated with a dashed yellow polygon.
The thickness of contour lines, when shown, are used to emphasise shape rather than surface brightness.
Panel (a): ALMA \coline\ line map. Contours are from the {\it HST} map of the H$\alpha$+[N II] emission line complex
at 6560 \AA, smoothed to match the angular resolution of the ALMA data. Panel (b): A map of the H$_{2}$ 1--0 S(1) line at 
2.12 $\mu$m from VLT/SINFONI. The contours of the ALMA \coline\ emission from Panel (a) are overlaid. Panel (c): A map of the ratio of 
F791W (optical) and F200N (near-infrared) images from {\it HST}, which emphasises dust absorption as dark features. 
The dust map is inaccurate at the nucleus (masked by a small white circle) because of excess near-infrared emission from hot dust around the 
AGN (see Section \ref{hstdata} for details). The blue contours show the shape of the ALMA 1 mm continuum, which traces
the bipolar radio jet in this AGN.
}
\label{contextfigure}
\end{figure*}

We used the K-band cube to measure the strength and kinematics of the H$_{2}$ 1--0 S(1) line at 2.12 $\mu$m,
modelled in each spaxel as a single gaussian with an underlying linear continuum.
A telluric residual from the reduction, masked appropriately in these fits, prevented
accurate line measurements in a few spaxels immediately around the continuum-bright nucleus. 
The apparent central hole in the H$_{2}$ S(1) maps of Figures \ref{contextfigure}b \& \ref{velocitymaps} 
are the consequence of this: the measurements in those spaxels have been excluded from any analysis.

From the J-band cube, we assessed the spatial structure of the [Fe II] 1.25 $\mu$m emission line
(Figure \ref{linemaps}). We fit this line using the same procedure as the 2.12 \mics\ H$_{2}$ S(1) line, 
but without the need to mask the central spaxels.

%

\subsection{Image registration and astrometry} 

The ALMA are astrometrically calibrated to the International Celestial Reference System (ICRS) with an accuracy $\approx 30$ mas. 
We have adopted the VLA radio core as the coordinates of
the nucleus (R.A.(J2000) $ = $ 5:52:11.379, Dec.(J2000) $ = $ -7:27:22.52), and verified that it lies within 30 mas of 
the peak of the unresolved nuclear core at 1 mm.

The {\it HST} images have accurate relative astrometry, but their absolute astrometry is noticeably incorrect. 
We derived a simple shift correction to the astrometric frame of the optical and NIR continuum images based on the difference between 
the centroidal positions and the absolute {\it GAIA} positions of two stars that lie within 20" of the galaxy centre. 
We visually verified that the peak of the NICMOS image lines up within 50 milliarcseconds 
of the radio nuclear position after we applied these astrometric corrections. 

We obtained NIR continuum maps directly from the SINFONI cubes, both of which show clear peaks. 
We registered the SINFONI cubes by tying the centroids of the continuum
maps to the radio nuclear position. The relative astrometry of SINFONI across its 
small field-of-view (FoV) is accurate enough for our purposes.

\subsection{{\it Spitzer}/IRS high-resolution spectroscopy} 

\begin{figure}[t]
\includegraphics[width=\columnwidth]{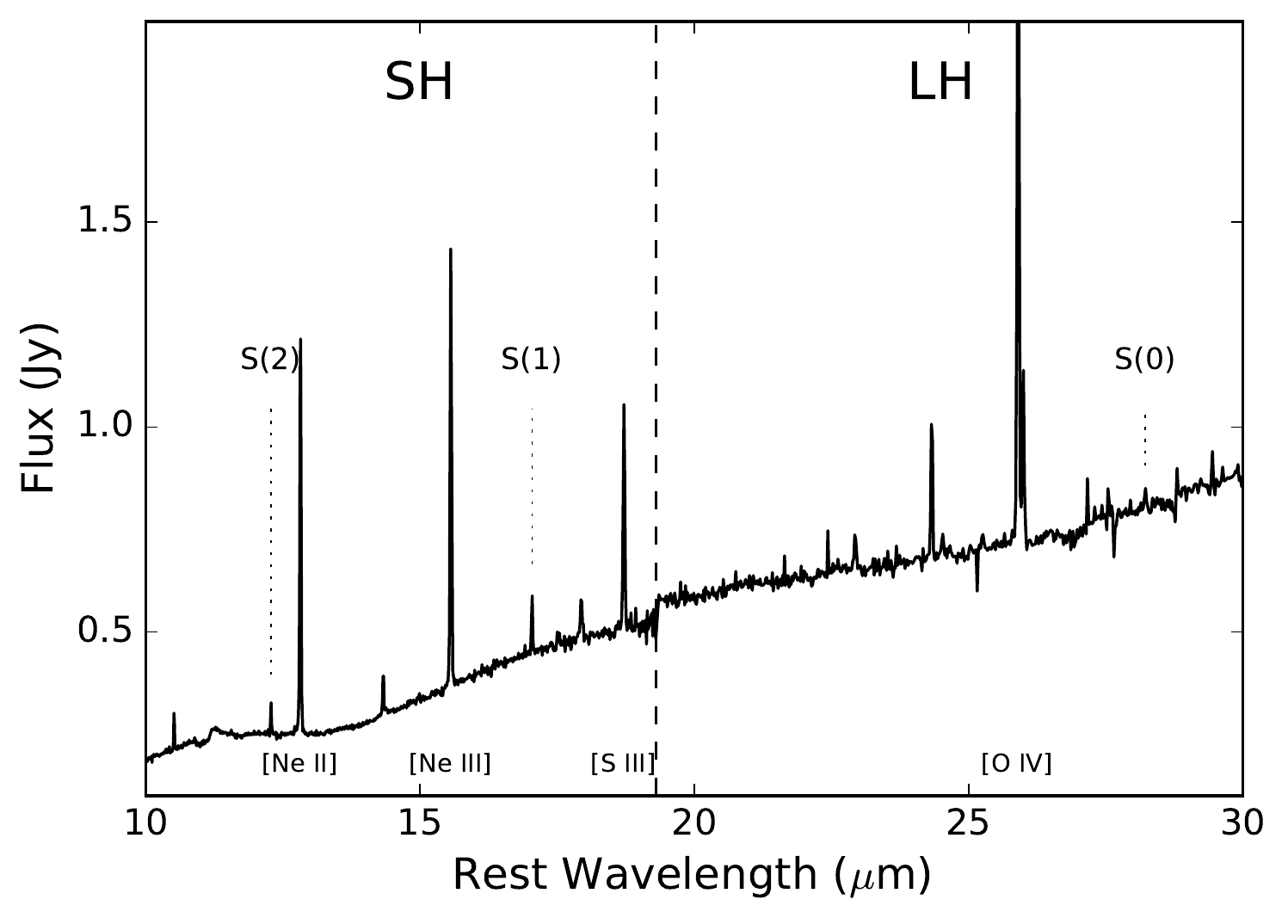}
\caption{The complete high-resolution {\it Spitzer}/IRS spectrum of NGC 2110 including both short (SH) and long (LH) spectral segments. 
Measurable H$_{2}$ 0--0 rotational emission lines are labelled with dotted line markers; prominent ionised gas emission lines are also identified.
}
\label{irsfig}
\end{figure}

We downloaded fully reduced, background-subtracted, optimally-extracted mid-infrared (MIR) 
spectra of NGC 2110 from the \href{http://cassis.sirtf.com/}{CASSIS} value-added database (Figure \ref{irsfig}).

We measured the fluxes of the MIR molecular hydrogen (H$_{2}$) 0--0 rotational lines at 
28.2 $\mu$m [S(0)], 17.0 $\mu$m [S(1)], and 12.3 $\mu$m [S(2)],
modeling each line as the combination of a single gaussian profile and an underlying linear continuum. 
The S(1) and S(2) lines are both well-detected with S/N $> 8$, while the S(0) line is marginally-detected with a S/N$\approx 2$.


The spectra from CASSIS are extracted following the spatial profile of the nuclear point source. 
The S(1,2) lines from the Short-High module (FWHM of $3\farcs5$ -- $6''$) are extracted over an area close to the SINFONI FoV, so 
aperture mismatch does not drastically affect our comparisons of these lines to the integrated H$_{2}$ 1--0 line emission 
(Section \ref{mir_modelling}). The PSF in the Long-High module at the S(0) line is considerably larger 
(FWHM of $9''$), and covers almost all of the detected CO emission seen in the ALMA maps.

\section{Direct evidence for AGN feedback on molecular gas in NGC 2110}

\subsection{A localised lack of cold molecular gas emission} \label{lacuna}

\begin{figure}[t]
\includegraphics[width=0.45\textwidth]{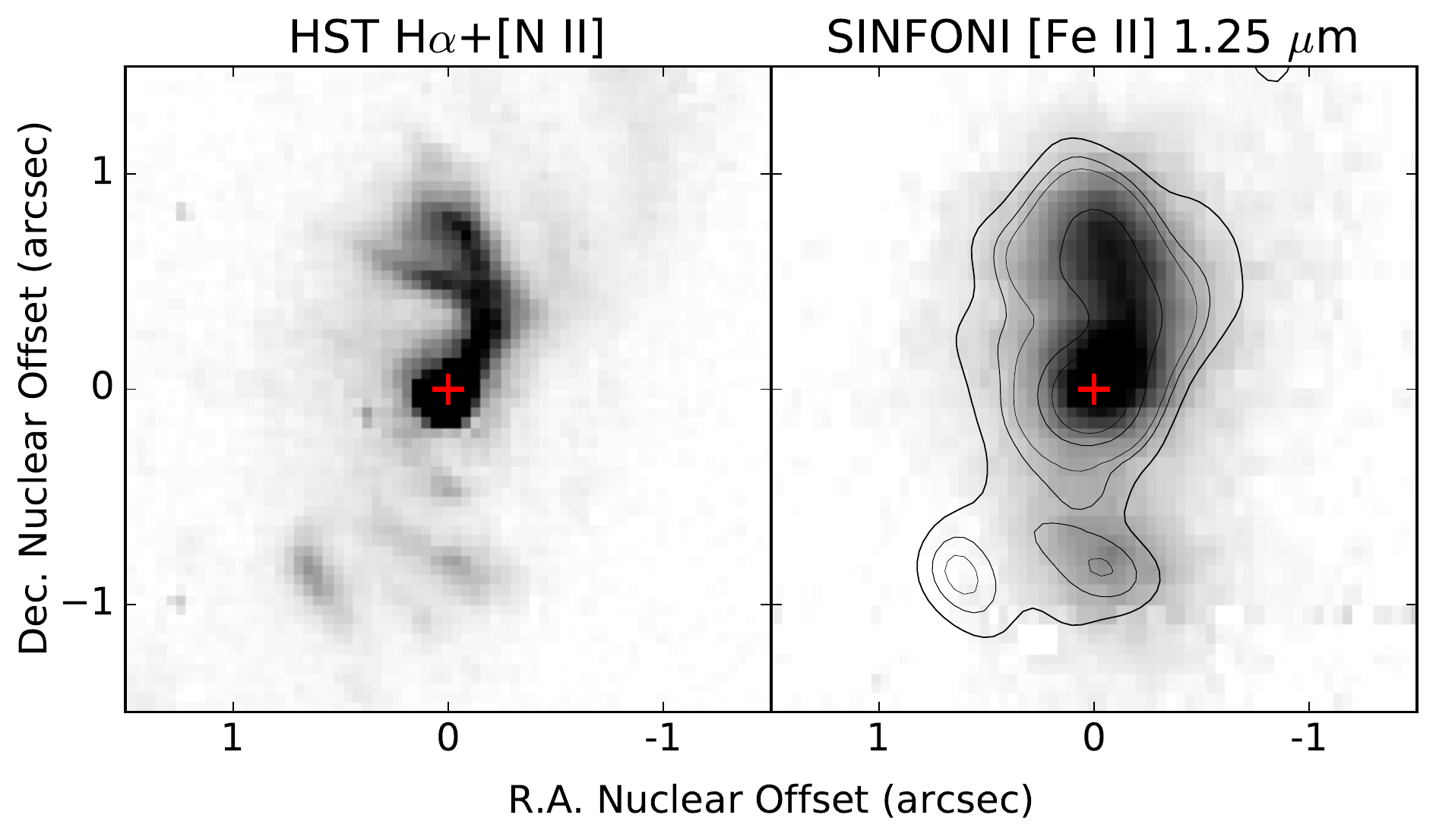}
\caption{ A comparison of emission line maps in the optical (left; the {\it  HST} H$\alpha$+[N II] complex at full resolution)
and the near-infrared (right; the [Fe II] 1.25 \mics\ line in the J-band from VLT/SINFONI). To highlight their similarity, 
we overlay the contours of the {\it HST} map in the right panel after matching it to the angular resolution of the SINFONI map.
The contour levels are unequally spaced; the lowest to highest contour levels 
span $9\times10^{-19}$ to $10^{-17}$ W m$^{-2}$ arcsec$^{-2}$.
The nucleus is marked with a cross in both panels; North is to the top and East is to the left. 
}
\label{linemaps}
\end{figure}

Figure \ref{contextfigure}a shows the ALMA \coline\ map in the centre of NGC 2110. 
This emission is distributed in an inhomogeneous spiral pattern suggestive of a circum-nuclear disc. 
Many of the bright arms of the CO disc are aligned with dark dust lanes seen 
in the {\it HST} color map (Figure \ref{contextfigure}c). 
For example, the brightest CO emission west of the nucleus is co-spatial with the dusty spiral arm on the near side of the galaxy
\citep[e.g., Section 6 of][]{rosario10}. 
Bordering this arm, one finds a conspicuous lack of CO emission in an extended linear structure passing through the 
nucleus at PA $\approx -25^{\circ}$, particularly within
a few arcseconds of the nucleus, where it bisects a region of high CO surface brightness,
but it also extends to the SE and NW of the nucleus. 
Henceforth, we use the term ``lacuna" to identify this feature.

The lacuna is well-resolved, and therefore unlikely to arise from \coline\ line absorption against the nuclear mm 
continuum in the galaxy \citep{tremblay16}, which is dominated by the well-known radio jet 
\citep[compare blue contour in Figure \ref{contextfigure}c to VLA 3.6 cm map in Figure 7 of][]{nagar99}.
NGC 2110 does not display a well-defined bi-symmetric pattern ($m=2$; grand design spiral or stellar bar),
so the separation of the two peaks of CO emission on either side of the lacuna cannot be easily attributed to stalling
at an Inner Lindblad Resonance, as has been noted in some barred galaxies \citep{kenney92}.


An examination of other excited ISM phases reveals a more intimate connection to the lacuna.
The 2.12 \mics\ H$_{2}$ 1--0 S(1) line, produced by hot excited molecular hydrogen, 
is located almost completely within the region (Figure \ref{contextfigure}b). A similar anticorrelation
between hot and cold molecular phases has been noted in other systems \citep[e.g.][]{davies04, davies14, mezcua15, espada17}.
Over the \coline\ map in Figure \ref{contextfigure}a, we have overlaid the contours 
from the H$\alpha$+[N II] emission line map. 
Studies have established this gas is ionised either by photoionisation from nuclear ultra-violet and X-ray light,
or via shocks from a fast wind with velocities of several 100 \kms\  \citep{ferruit99, rosario10, schnorr14}.
The narrow bi-polar shape may be due to the anisotropic illumination of the circum-nuclear disk by the AGN 
\citep[e.g., Figure 7 of][]{rosario10}.  

Figure \ref{contextfigure}a reveals a close spatial association between the CO lacuna and the AGN-ionised emission line gas. 
The two structures are highly co-spatial, and the CO emission is noticeably weaker along the axis of the ionised gas.
Within an arcsecond of the nucleus, the inner edges of the lacuna are defined by bright CO features which mirror the outer 
edges of the emission line region. 
 
The cold, dusty gas that produces \coline\ could potentially shape the observed optical emission line structure through
selective extinction, resulting in an apparent anti-correlation between the two phases. 
We test this by examining a map of the 1.25 $\mu$m [Fe II] line, which is also excited by the AGN, but
is less extinguished by dust than H$\alpha$+[N II]. The similarity of the two maps 
(Figure \ref{linemaps}) confirms that the intrinsic structure of the AGN-ionised region is 
accurately represented by the contours in Figure \ref{contextfigure}a. NIR hydrogen recombination lines,
such as Br$\gamma$ at 2.17 $\mu$m, also share the same basic size and structure \citep{diniz15}.

Interestingly, the {\it HST} color map (Figure \ref{contextfigure}c) also reveals considerable dusty 
gas within the lacuna which is not visible in \coline. At larger nuclear distances, the ionised gas traces spiral features 
visible in the {\it HST} dust map, yet the CO emission here also remains weak. 

\subsection{Associated enhancement in warm molecular gas emission} \label{mir_modelling}

\begin{figure}[t]
\includegraphics[width=\columnwidth]{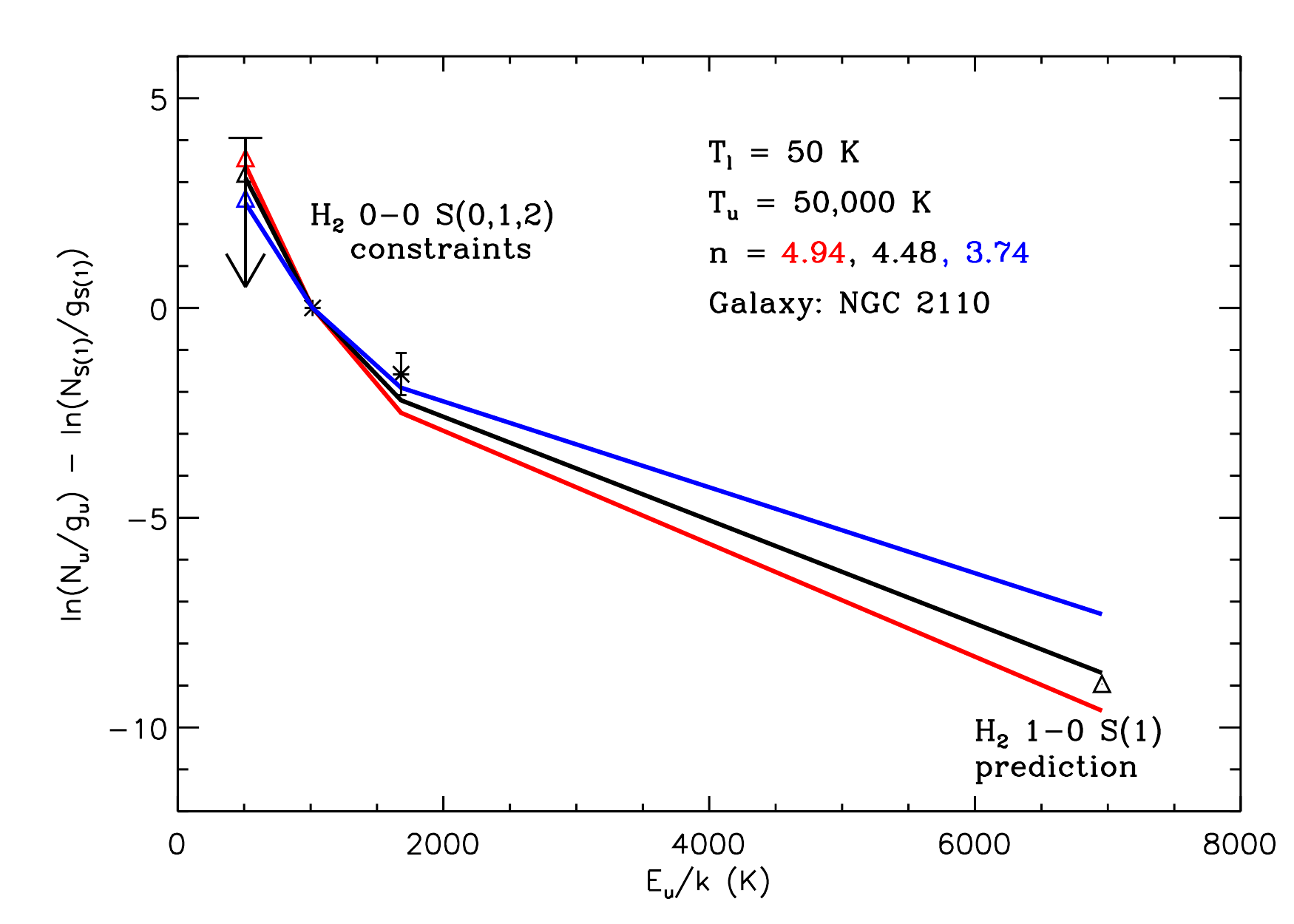}
\caption{Excitation diagram of H$_{2}$ showing our power-law fit to the mid-infrared 0--0 rotational line strengths and the extrapolation
of the models to the 1--0 S(1) line at 2.12 $\mu$m. Colored lines correspond to models with different power-law indices ($n$) as shown in the key. The downward arrow is the 3$\sigma$ upper limit on the 0--0 S(0) constraint. See Section \ref{mir_modelling} 
for more details.
}
\label{h2model}
\end{figure}

Fundamental insight into the nature of the lacuna comes from the modeling of the molecular line sequence
of warm H$_{2}$ from {\it Spitzer}/IRS spectroscopy. 

In the complex environment of a galaxy nucleus, a few discrete temperature components do not adequately describe the excitation of H$_{2}$ gas. 
Adopting the approach of \citet{togi16}, we assume a uniform power law model of the distribution of H$_{2}$ temperatures, with the form
dN $\propto$ T$^{-n}$ dT, where dN is the column density of molecules in the temperature range T---T+dT.
The two free parameters are the power law index $n$ and the lower temperature T$_{\ell}$ respectively. 
The upper temperature of the distribution is fixed at $50000$ K, though realistically the fraction of molecular mass with temperatures $>$ a few 1000 K is negligible.

Figure \ref{h2model} shows an H$_{2}$ excitation diagram that illustrates the constraints offered by the measured
H$_{2}$ rotational lines (including limits), and the associated uncertainties on the power law index. 
In the diagram, we plot the column density of molecules 
populated by the upper level of a transition (N$_{\rm u}$) divided by its statistical weight (g$_{\rm u}$), against the energy level of the
transition (E$_{\rm u}$). We follow the custom of normalising the excitation to the 0--0 S(1) line \citep{togi16}. 
Extrapolating the power-law model to temperatures $>1000$ K gives an estimate of the flux of the 1--0 S(1) line at 2.12 $\mu$m,
which also serves as a constraint. The hot H$_{2}$ gas that emits this line is a very small fraction (typically $<0.1$\%) of the 
total molecular mass (M$_{\rm mol}$), therefore this extrapolation is strictly contingent on the continuity of the 
power law distribution of temperatures beyond several 100 K. The similarity of the rotational and vibrational temperatures 
derived from NIR H$_{2}$ lines implies that even the hot molecular material is in thermal equilibrium \citep{diniz15}, 
lending some support to this assumption.

Fixing the model to the formally measured flux of the S(0) line, we obtain $n=4.48$ (black line in Figure \ref{h2model}). 
This value is towards the upper end of the range found among star-forming galaxies in the Spitzer Infrared Nearby Galaxies Survey \citep{togi16}. Such a shallow temperature distribution arises from a higher mass fraction of warm H$_{2}$ than typically found in 
galaxy environments. Considering temperatures as low as $ 50$ K to include the cold component that emits \coline, 
we calculate M$_{\rm mol} = 2.5\times10^{8}$ \msun, including the contribution of helium and heavier elements

From Figure \ref{h2model}, it is clear that the adopted strength of the 0--0 S(0) line strongly 
influences the determination of $n$ and therefore the final estimate of M$_{\rm mol}$. A nominal uncertainty of 0.3 dex for the line flux
implies a mass in the range of $0.9$--$4.6\times10^{8}$ \msun. The estimated mass is correlated with $n$: 
a larger proportion of molecules at high temperatures (lower $n$) results in a lower estimate of M$_{\rm mol}$. 
In addition, the S(0) constraint should be formally considered an upper limit since the aperture used for the measurement of this line 
covers a substantially larger area than the lacuna itself; the arrow in Figure \ref{h2model} shows the equivalent 3$\sigma$
limit on the line. Therefore, the molecular mass associated with the lacuna could be even smaller than the range calculated above.

However, in this regard, the 2.12 $\mu$m 1--0 S(1) provides a measure of discriminatory power. Figure \ref{h2model} shows that
a single power-law model tied to the formal flux of the S(0) line can reproduce the fluxes measured in all four H$_{2}$ lines quite well.
This suggests that the S(0) line flux is not very extended, but mostly concentrated within the lacuna and its immediate surroundings.

We can also estimate the total molecular gas mass directly from the 
\coline\ line over the same region ($F_{\rm CO} = 13$ Jy \kms from a 4" circular aperture centered on the nucleus).
Following \citet{solomon05}: 

\begin{equation}
{\rm M}_{\rm mol, CO} = \frac{3.25\times10^{7} \, R_{12} \, \alpha_{\rm CO}  F_{\rm CO} \,D_{L}^{2} }{(1+z) \times (230.54\, {\rm GHz})^{2}}  \; \; M_{\odot}  \label{eq1}
\end{equation} 

\noindent assuming a certain CO-to-H$_{2}$ conversion factor ($\alpha_{\rm CO}$) and a CO 1$\rightarrow$0
to \coline\ brightness temperature ratio ($R_{12}$).
Taking $R_{12} = 1.4$ and $\alpha_{\rm CO}$ in the range of 1.5 -- 3, consistent with observations of the centers of 
nearby galaxies \citep{sandstrom13}, we obtain M$_{\rm mol, CO}\approx2 - 4 \times10^{7}$ \msun. This is 
considerably lower than M$_{\rm mol}$ estimated from the MIR H$_{2}$ lines.

We postulate that the molecular mass invisible in \coline\ has been
heated beyond the temperature at which it efficiently emits low-order CO lines, and
this material is concentrated in the lacuna and shares the 
spatial distribution of the NIR H$_{2}$ S(1) line, a circular region of 0.34 kpc$^{2}$.
From the difference ${\rm M}_{\rm mol} - {\rm M}_{\rm mol, CO} \approx 2.2 \times10^{8}$ \msun, 
we infer a molecular gas surface density of 650 \msun\ pc$^{-2}$. 
If we treat the S(0) line as a formal limit, and take ${\rm M}_{\rm mol} \approx 9 \times10^{7}$ \msun,
at the low end of the estimated range, the surface density drops to 180 \msun\ pc$^{-2}$.
These calculations may be compared to the 
molecular gas surface density of $200$-- $350$ \msun\ pc$^{-2}$ inferred using Equation \ref{eq1}
in the bright CO knots around the edges of the lacuna. 

The similarity of these two estimates, certainly within the 
systematic uncertainties of our modeling, suggests that the central molecular disk extends into the lacuna, 
despite its apparent CO deficiency.

\begin{figure}[t]
\includegraphics[width=\columnwidth]{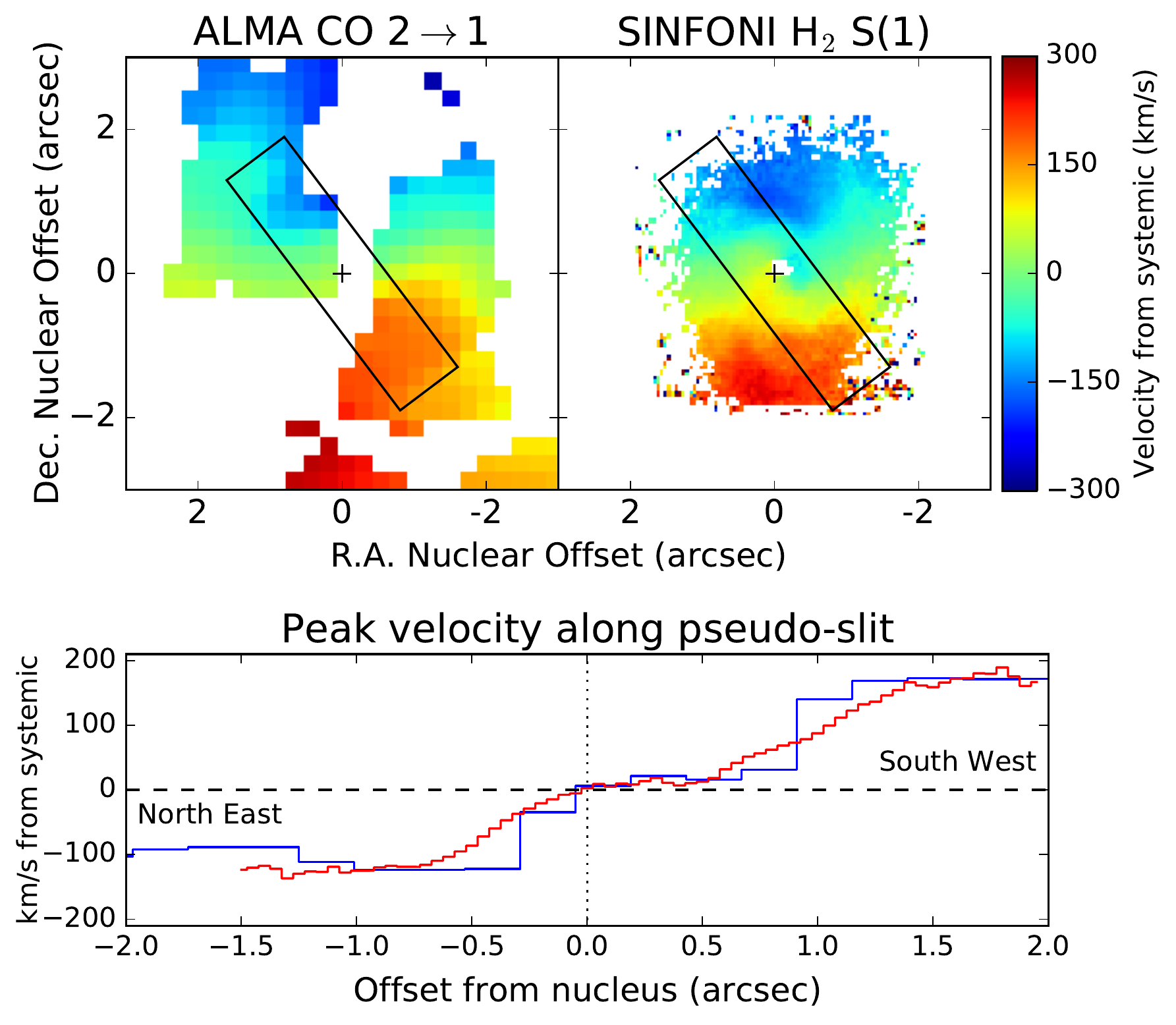}
\caption{Top: Peak systemic velocity offset of the \coline\ line (left) and the 2.12 $\mu$m H$_{2}$ 1--0 S(1) line (right) 
from ALMA and SINFONI datacubes. Bottom: Velocity curves extracted from the same simulated long-slit aperture
(the black rectangle in both top panels) from the CO (blue) and H$_{2}$ (red) datasets. Velocity errors, 
in the range of 2--10 km s$^{-1}$, are excluded for visual purposes.
While the distribution of the cold and warm molecular gas are different, they share the same velocity field.}
\label{velocitymaps}
\end{figure}

Additional support comes from the comparison of the two-dimensional velocity fields of the \coline\ line and the H$_{2}$ 1--0 S(1) line
(Figure \ref{velocitymaps}). Despite differences in the temperature and excitation of these molecular species, 
both lines independently trace an inclined rotating disc with the same 
strong kinematic asymmetry and non-circular motions as have been found
previously from ionised gas studies \citep{gonzalez02, ferruit04, schnorr14, diniz15}.
From the continuity in the rotation fields of the cold and warm molecular gas, we infer that these two phases 
are connected and share the same circum-nuclear dynamics.

\subsection{AGN feedback suppresses \coline\ emission}

Based on the evidence developed above, the most likely explanation for the CO lacuna is that the
energy liberated by the central AGN is directly influencing the molecular gas within the region
and actively suppressing the emission of the \coline\ line. 
This ``AGN feedback''  can proceed in two possible ways.
Strong far-ultraviolet and X-ray radiation from the AGN penetrates past the ionised outer layers of dense 
clouds in the lacuna, heating the molecular gas within. This alters its chemistry, while photo-excitation and dissociation of CO 
suppresses the emission of the \coline\ line. Alternatively, slow shocks ($\lesssim 50$ \kms)
arising from the interaction between molecular gas and an AGN wind or radio jet depletes CO while boosting
the emission of the warm molecular hydrogen lines. The molecular gas diagnostics currently available 
do not strongly discriminate between these mechanisms. The relative strengths of the 
rotational and ro-vibrational H$_{2}$ lines \citep{ardila05, diniz15} can be achieved with models that feature either 
shock and X-ray excitation \citep{maloney96, rigopoulou02, flower10}.  

However, the entire optical and NIR line spectrum of NGC 2110, including the warm molecular lines, can be self-consistently reproduced 
by photoionisation of metal-rich dusty gas ($\approx 2\times$ the solar metal abundance) by a nuclear source with the known 
power of the AGN \citep{rosario10, dors12}. In light of this, the close relationship between the structure of the ionised gas, 
the warm H$_{2}$, and the CO lacuna supports radiative feedback as the principal cause for the transformation of the molecular gas.
 
Regardless of the primary process that suppresses the CO emission, our study concludes that an 
AGN can directly influence the local emissive and thermal properties of
circum-nuclear molecular gas. This is the first time such a strong association has been noted with such clarity.
\citet{querejeta16} have reported a similar connection in M51, though the conclusion is limited
by the resolution of their CO data. In NGC 1068, the archetypical local Seyfert, the CO emission appears to be 
decoupled from its well-known ionisation cone \citep[e.g., Figure 6 in][]{garciaburillo14}.  NGC 5643 may show
some evidence in an extended arm of CO emission intersecting the ionisation cone \citep[e.g., Figure 1 in][]{alonsoherrero18}.

An important corollary worth highlighting is that such interaction could be quite localised. There is considerable
molecular material in the vicinity of the nucleus of NGC 2110 which remains free of any obvious nuclear impact. 
Even within the lacuna, we find that molecular material can remain resilient to the mechanical effects
of radiation pressure or AGN winds. This has important implications for the role of AGN feedback
in regulating and suppressing star-formation in galaxies. NGC 2110 will serve as a valuable laboratory to explore 
this key process that underpins our modern theoretical view of galaxy evolution.


\acknowledgments
DR acknowledges the support of the Science and Technology Facilities Council (STFC) through grant ST/P000541/1.
ALMA is a partnership of ESO (representing its member states), NSF (USA) and NINS (Japan), 
together with NRC (Canada), MOST and ASIAA (Taiwan), and KASI (Republic of Korea), in cooperation with the Republic of Chile.
Based on observations collected at the European Organisation for Astronomical Research in the Southern Hemisphere;
with the NASA/ESA Hubble Space Telescope, obtained from the data archive at the Space Telescope Science Institute, which is
operated by the Association of Universities for Research in Astronomy, Inc. under NASA contract NAS 5-26555; 
and with the Spitzer Space Telescope, which is operated by the Jet Propulsion Laboratory, California Institute of Technology under a contract with NASA.

\bibliography{NGC2110-ALMA}{}

\begin{thebibliography}{}
\expandafter\ifx\csname natexlab\endcsname\relax\def\natexlab#1{#1}\fi
\providecommand{\url}[1]{\href{#1}{#1}}

\bibitem[{{Alexander} \& {Hickox}(2012)}]{alexander12}
{Alexander}, D.~M., \& {Hickox}, R.~C. 2012, \nar, 56, 93

\bibitem[{{Alonso-Herrero} {et~al.}(2018){Alonso-Herrero}, {Pereira-Santaella},
  {Garc{\'{\i}}a-Burillo}, {Davies}, {Combes}, {Asmus}, {Bunker},
  {D{\'{\i}}az-Santos}, {Gandhi}, {Gonz{\'a}lez-Mart{\'{\i}}n},
  {Hern{\'a}n-Caballero}, {Hicks}, {H{\"o}nig}, {Labiano}, {Levenson},
  {Packham}, {Ramos Almeida}, {Ricci}, {Rigopoulou}, {Rosario}, {Sani}, \&
  {Ward}}]{alonsoherrero18}
{Alonso-Herrero}, A., {Pereira-Santaella}, M., {Garc{\'{\i}}a-Burillo}, S.,
  {et~al.} 2018, \apj, 859, 144

\bibitem[{{Bigiel} {et~al.}(2008){Bigiel}, {Leroy}, {Walter}, {Brinks}, {de
  Blok}, {Madore}, \& {Thornley}}]{bigiel08}
{Bigiel}, F., {Leroy}, A., {Walter}, F., {et~al.} 2008, Astron. J., 136, 2846

\bibitem[{{Burtscher} {et~al.}(2015){Burtscher}, {Orban de Xivry}, {Davies},
  {Janssen}, {Lutz}, {Rosario}, {Contursi}, {Genzel}, {Graci{\'a}-Carpio},
  {Lin}, {Schnorr-M{\"u}ller}, {Sternberg}, {Sturm}, \&
  {Tacconi}}]{burtscher15}
{Burtscher}, L., {Orban de Xivry}, G., {Davies}, R.~I., {et~al.} 2015, \aap,
  578, A47

\bibitem[{{Davies} {et~al.}(2004){Davies}, {Tacconi}, \& {Genzel}}]{davies04}
{Davies}, R.~I., {Tacconi}, L.~J., \& {Genzel}, R. 2004, \apj, 602, 148

\bibitem[{{Davies} {et~al.}(2014){Davies}, {Maciejewski}, {Hicks}, {Emsellem},
  {Erwin}, {Burtscher}, {Dumas}, {Lin}, {Malkan}, {M{\"u}ller-S{\'a}nchez},
  {Orban de Xivry}, {Rosario}, {Schnorr-M{\"u}ller}, \& {Tran}}]{davies14}
{Davies}, R.~I., {Maciejewski}, W., {Hicks}, E.~K.~S., {et~al.} 2014, \apj,
  792, 101

\bibitem[{{Diniz} {et~al.}(2015){Diniz}, {Riffel}, {Storchi-Bergmann}, \&
  {Winge}}]{diniz15}
{Diniz}, M.~R., {Riffel}, R.~A., {Storchi-Bergmann}, T., \& {Winge}, C. 2015,
  \mnras, 453, 1727

\bibitem[{{Dors} {et~al.}(2012){Dors}, {Riffel}, {Cardaci}, {H{\"a}gele},
  {Krabbe}, {P{\'e}rez-Montero}, \& {Rodrigues}}]{dors12}
{Dors}, Jr., O.~L., {Riffel}, R.~A., {Cardaci}, M.~V., {et~al.} 2012, \mnras,
  422, 252

\bibitem[{{Espada} {et~al.}(2017){Espada}, {Matsushita}, {Miura}, {Israel},
  {Neumayer}, {Martin}, {Henkel}, {Izumi}, {Iono}, {Aalto}, {Ott}, {Peck},
  {Quillen}, \& {Kohno}}]{espada17}
{Espada}, D., {Matsushita}, S., {Miura}, R.~E., {et~al.} 2017, \apj, 843, 136

\bibitem[{{Ferruit} {et~al.}(2004){Ferruit}, {Mundell}, {Nagar}, {Emsellem},
  {P{\'e}contal}, {Wilson}, \& {Schinnerer}}]{ferruit04}
{Ferruit}, P., {Mundell}, C.~G., {Nagar}, N.~M., {et~al.} 2004, Mon. Not. Royal
  Astron. Soc., 352, 1180

\bibitem[{{Ferruit} {et~al.}(1999){Ferruit}, {Wilson}, {Whittle}, {Simpson},
  {Mulchaey}, \& {Ferland}}]{ferruit99}
{Ferruit}, P., {Wilson}, A.~S., {Whittle}, M., {et~al.} 1999, \apj, 523, 147

\bibitem[{{Flower} \& {Pineau Des For{\^e}ts}(2010)}]{flower10}
{Flower}, D.~R., \& {Pineau Des For{\^e}ts}, G. 2010, \mnras, 406, 1745

\bibitem[{{Garc{\'{\i}}a-Burillo} {et~al.}(2014){Garc{\'{\i}}a-Burillo},
  {Combes}, {Usero}, {Aalto}, {Krips}, {Viti}, {Alonso-Herrero}, {Hunt},
  {Schinnerer}, {Baker}, {Boone}, {Casasola}, {Colina}, {Costagliola},
  {Eckart}, {Fuente}, {Henkel}, {Labiano}, {Mart{\'{\i}}n}, {M{\'a}rquez},
  {Muller}, {Planesas}, {Ramos Almeida}, {Spaans}, {Tacconi}, \& {van der
  Werf}}]{garciaburillo14}
{Garc{\'{\i}}a-Burillo}, S., {Combes}, F., {Usero}, A., {et~al.} 2014, Astron.
  Astrophys., 567, A125

\bibitem[{{Gonz{\'a}lez Delgado} {et~al.}(2002){Gonz{\'a}lez Delgado},
  {Arribas}, {P{\'e}rez}, \& {Heckman}}]{gonzalez02}
{Gonz{\'a}lez Delgado}, R.~M., {Arribas}, S., {P{\'e}rez}, E., \& {Heckman}, T.
  2002, \apj, 579, 188

\bibitem[{{Heckman} \& {Best}(2014)}]{heckman14}
{Heckman}, T.~M., \& {Best}, P.~N. 2014, Ann. Rev. Astro. Astrophy., 52, 589

\bibitem[{{Imanishi} {et~al.}(2016){Imanishi}, {Nakanishi}, \&
  {Izumi}}]{imanishi16}
{Imanishi}, M., {Nakanishi}, K., \& {Izumi}, T. 2016, \aj, 152, 218

\bibitem[{{Izumi} {et~al.}(2013){Izumi}, {Kohno}, {Mart{\'{\i}}n}, {Espada},
  {Harada}, {Matsushita}, {Hsieh}, {Turner}, {Meier}, {Schinnerer}, {Imanishi},
  {Tamura}, {Curran}, {Doi}, {Fathi}, {Krips}, {Lundgren}, {Nakai}, {Nakajima},
  {Regan}, {Sheth}, {Takano}, {Taniguchi}, {Terashima}, {Tosaki}, \&
  {Wiklind}}]{izumi13}
{Izumi}, T., {Kohno}, K., {Mart{\'{\i}}n}, S., {et~al.} 2013, \pasj, 65, 100

\bibitem[{{Izumi} {et~al.}(2016){Izumi}, {Kohno}, {Aalto}, {Espada}, {Fathi},
  {Harada}, {Hatsukade}, {Hsieh}, {Imanishi}, {Krips}, {Mart{\'{\i}}n},
  {Matsushita}, {Meier}, {Nakai}, {Nakanishi}, {Schinnerer}, {Sheth},
  {Terashima}, \& {Turner}}]{izumi16}
{Izumi}, T., {Kohno}, K., {Aalto}, S., {et~al.} 2016, \apj, 818, 42

\bibitem[{{Kenney} {et~al.}(1992){Kenney}, {Wilson}, {Scoville}, {Devereux}, \&
  {Young}}]{kenney92}
{Kenney}, J.~D.~P., {Wilson}, C.~D., {Scoville}, N.~Z., {Devereux}, N.~A., \&
  {Young}, J.~S. 1992, \apjl, 395, L79

\bibitem[{{Kennicutt}(1989)}]{kennicutt89}
{Kennicutt}, Jr., R.~C. 1989, Astrophys. J., 344, 685

\bibitem[{{Kohno} {et~al.}(2003){Kohno}, {Ishizuki}, {Matsushita},
  {Vila-Vilar{\'o}}, \& {Kawabe}}]{kohno03}
{Kohno}, K., {Ishizuki}, S., {Matsushita}, S., {Vila-Vilar{\'o}}, B., \&
  {Kawabe}, R. 2003, \pasj, 55, L1

\bibitem[{{Kohno} {et~al.}(2008){Kohno}, {Nakanishi}, {Tosaki}, {Muraoka},
  {Miura}, {Ezawa}, \& {Kawabe}}]{kohno08}
{Kohno}, K., {Nakanishi}, K., {Tosaki}, T., {et~al.} 2008, \apss, 313, 279

\bibitem[{{Kormendy} \& {Ho}(2013)}]{kormendy13}
{Kormendy}, J., \& {Ho}, L.~C. 2013, \araa, 51, 511

\bibitem[{{Krips} {et~al.}(2008){Krips}, {Neri}, {Garc{\'{\i}}a-Burillo},
  {Mart{\'{\i}}n}, {Combes}, {Graci{\'a}-Carpio}, \& {Eckart}}]{krips08}
{Krips}, M., {Neri}, R., {Garc{\'{\i}}a-Burillo}, S., {et~al.} 2008, Astrophys.
  J., 677, 262

\bibitem[{{Krips} {et~al.}(2011){Krips}, {Mart{\'{\i}}n}, {Eckart}, {Neri},
  {Garc{\'{\i}}a-Burillo}, {Matsushita}, {Peck}, {Stoklasov{\'a}}, {Petitpas},
  {Usero}, {Combes}, {Schinnerer}, {Humphreys}, \& {Baker}}]{krips11}
{Krips}, M., {Mart{\'{\i}}n}, S., {Eckart}, A., {et~al.} 2011, \apj, 736, 37

\bibitem[{{Maloney} {et~al.}(1996){Maloney}, {Hollenbach}, \&
  {Tielens}}]{maloney96}
{Maloney}, P.~R., {Hollenbach}, D.~J., \& {Tielens}, A.~G.~G.~M. 1996, \apj,
  466, 561

\bibitem[{{Mezcua} {et~al.}(2015){Mezcua}, {Prieto}, {Fern{\'a}ndez-Ontiveros},
  {Tristram}, {Neumayer}, \& {Kotilainen}}]{mezcua15}
{Mezcua}, M., {Prieto}, M.~A., {Fern{\'a}ndez-Ontiveros}, J.~A., {et~al.} 2015,
  \mnras, 452, 4128

\bibitem[{{Nagar} {et~al.}(1999){Nagar}, {Wilson}, {Mulchaey}, \&
  {Gallimore}}]{nagar99}
{Nagar}, N.~M., {Wilson}, A.~S., {Mulchaey}, J.~S., \& {Gallimore}, J.~F. 1999,
  \apjs, 120, 209

\bibitem[{{Querejeta} {et~al.}(2016){Querejeta}, {Schinnerer},
  {Garc{\'{\i}}a-Burillo}, {Bigiel}, {Blanc}, {Colombo}, {Hughes}, {Kreckel},
  {Leroy}, {Meidt}, {Meier}, {Pety}, \& {Sliwa}}]{querejeta16}
{Querejeta}, M., {Schinnerer}, E., {Garc{\'{\i}}a-Burillo}, S., {et~al.} 2016,
  Astron. Astrophys., 593, A118

\bibitem[{{Rigopoulou} {et~al.}(2002){Rigopoulou}, {Kunze}, {Lutz}, {Genzel},
  \& {Moorwood}}]{rigopoulou02}
{Rigopoulou}, D., {Kunze}, D., {Lutz}, D., {Genzel}, R., \& {Moorwood},
  A.~F.~M. 2002, \aap, 389, 374

\bibitem[{{Rodr{\'{\i}}guez-Ardila} {et~al.}(2005){Rodr{\'{\i}}guez-Ardila},
  {Riffel}, \& {Pastoriza}}]{ardila05}
{Rodr{\'{\i}}guez-Ardila}, A., {Riffel}, R., \& {Pastoriza}, M.~G. 2005,
  \mnras, 364, 1041

\bibitem[{{Rosario} {et~al.}(2010){Rosario}, {Whittle}, {Nelson}, \&
  {Wilson}}]{rosario10}
{Rosario}, D.~J., {Whittle}, M., {Nelson}, C.~H., \& {Wilson}, A.~S. 2010, Mon.
  Not. Royal Astron. Soc., 408, 565

\bibitem[{{Sandstrom} {et~al.}(2013){Sandstrom}, {Leroy}, {Walter}, {Bolatto},
  {Croxall}, {Draine}, {Wilson}, {Wolfire}, {Calzetti}, {Kennicutt}, {Aniano},
  {Donovan Meyer}, {Usero}, {Bigiel}, {Brinks}, {de Blok}, {Crocker}, {Dale},
  {Engelbracht}, {Galametz}, {Groves}, {Hunt}, {Koda}, {Kreckel}, {Linz},
  {Meidt}, {Pellegrini}, {Rix}, {Roussel}, {Schinnerer}, {Schruba}, {Schuster},
  {Skibba}, {van der Laan}, {Appleton}, {Armus}, {Brandl}, {Gordon}, {Hinz},
  {Krause}, {Montiel}, {Sauvage}, {Schmiedeke}, {Smith}, \&
  {Vigroux}}]{sandstrom13}
{Sandstrom}, K.~M., {Leroy}, A.~K., {Walter}, F., {et~al.} 2013, \apj, 777, 5

\bibitem[{{Schnorr-M{\"u}ller} {et~al.}(2014){Schnorr-M{\"u}ller},
  {Storchi-Bergmann}, {Nagar}, {Robinson}, {Lena}, {Riffel}, \&
  {Couto}}]{schnorr14}
{Schnorr-M{\"u}ller}, A., {Storchi-Bergmann}, T., {Nagar}, N.~M., {et~al.}
  2014, \mnras, 437, 1708

\bibitem[{{Solomon} \& {Vanden Bout}(2005)}]{solomon05}
{Solomon}, P.~M., \& {Vanden Bout}, P.~A. 2005, \araa, 43, 677

\bibitem[{{Sternberg} {et~al.}(1994){Sternberg}, {Genzel}, \&
  {Tacconi}}]{sternberg94}
{Sternberg}, A., {Genzel}, R., \& {Tacconi}, L. 1994, Astrophys. J. Lett., 436,
  L131

\bibitem[{{Togi} \& {Smith}(2016)}]{togi16}
{Togi}, A., \& {Smith}, J.~D.~T. 2016, \apj, 830, 18

\bibitem[{{Tremblay} {et~al.}(2016){Tremblay}, {Oonk}, {Combes}, {Salom{\'e}},
  {O'Dea}, {Baum}, {Voit}, {Donahue}, {McNamara}, {Davis}, {McDonald}, {Edge},
  {Clarke}, {Galv{\'a}n-Madrid}, {Bremer}, {Edwards}, {Fabian}, {Hamer}, {Li},
  {Maury}, {Russell}, {Quillen}, {Urry}, {Sanders}, \& {Wise}}]{tremblay16}
{Tremblay}, G.~R., {Oonk}, J.~B.~R., {Combes}, F., {et~al.} 2016, \nat, 534,
  218

\bibitem[{{Usero} {et~al.}(2004){Usero}, {Garc{\'{\i}}a-Burillo}, {Fuente},
  {Mart{\'{\i}}n-Pintado}, \& {Rodr{\'{\i}}guez-Fern{\'a}ndez}}]{usero04}
{Usero}, A., {Garc{\'{\i}}a-Burillo}, S., {Fuente}, A.,
  {Mart{\'{\i}}n-Pintado}, J., \& {Rodr{\'{\i}}guez-Fern{\'a}ndez}, N.~J. 2004,
  Astron. Astrophys., 419, 897

\end{thebibliography}

\end{document}